\documentclass[runningheads]{llncs}
\usepackage{graphicx}
\usepackage{booktabs}
\usepackage{microtype}
\usepackage[misc]{ifsym}
\usepackage{orcidlink}

\begin{document}
\title{Cardiac Segmentation using Transfer Learning under Respiratory Motion Artifacts 
}
\titlerunning{Cardiac Segmentation Under Respiratory Artifacts.}
%
\author{Carles Garcia-Cabrera\inst{1}\Letter \orcidlink{0000-0001-8139-9647}\and
Eric Arazo\inst{1} \orcidlink{0000-0001-9769-3592} \and
Kathleen M. Curran\inst{2}\orcidlink{0000-0003-0095-9337} \and
Noel E. O'Connor\inst{1} \orcidlink{0000-0002-4033-9135} \and
Kevin McGuinness\inst{1} \orcidlink{0000-0003-1336-6477}
}
%

\authorrunning{C. Garcia-Cabrera et al.}
%

\institute{Dublin City University, Dublin, Ireland \\ \email{carles.garciacabrera6@mail.dcu.ie}\and
University College Dublin, Dublin, Ireland}
\maketitle              
\begin{abstract}
Methods that are resilient to artifacts in the cardiac magnetic resonance imaging (MRI) while performing ventricle segmentation, are crucial for ensuring quality in structural and functional analysis of those tissues. While there has been significant efforts on improving the quality of the algorithms, few works have tackled the harm that the artifacts generate in the predictions. In this work, we study fine tuning of pretrained networks to improve the resilience of previous methods to these artifacts. In our proposed method, we adopted the extensive usage of data augmentations that mimic those artifacts. The results significantly improved the baseline segmentations (up to 0.06 Dice score, and 4mm Hausdorff distance improvement). 

\end{abstract}
\section{Introduction}
Cardiac-related diseases (CVDs) are the most prominent cause of death globally \cite{World}. To accelerate, cheapen and enhance the quality of the diagnostics and treatment of the patient with cardiovascular diseases, the related techniques and technologies should be accurate and efficient. Non-invasive imaging modalities such as cardiac magnetic resonance imaging (CMRI) are particularly useful in the clinical assessment due to being capable of providing detailed data from patients. Extracting morphological and functional information from these data is tedious and intensive and can lead to observer bias \cite{Petitjean_Dacher_2011}.

The automation of these tasks has attracted the attention of scientists due to its high impact on daily clinical workflows. In the last few years, the innovation of deep learning techniques, and in particular convolutional neural networks, have brought more interest in the topic and have demonstrated great potential~\cite{Isensee_Jaeger_Kohl_Petersen_Maier-Hein_2021,rueckert,imvip}. While recent efforts address approaches to resolve the quality of the segmentation of all the chambers within the heart (right ventricle, left ventricle, and myocardium) and multiple views of the tissue (long-axis and short-axis) \cite{mms}; sub-optimal segmentations, e.g. those affected by respiratory artifacts, are still under-explored.

To address this, in this work, we propose a method to benefit from pretrained models that are publicly available. Using the pretrained weights not only hasted the training process, but also enhanced the predictions of the network. In particular, the results were substantially better for the right ventricle (RV) and the myocardium (MYO), leading to an increase in DICE and a reduction in Hausdorff. Results on the left ventricle (LV), however, remained unchanged. This work was done in the context of the Extreme Cardiac MRI Analysis Challenge under Respiratory Motion (CMRxMotion), where we focused on Task 2 (the segmentation task).

\begin{figure}[ht!]
    \includegraphics[width=\textwidth]{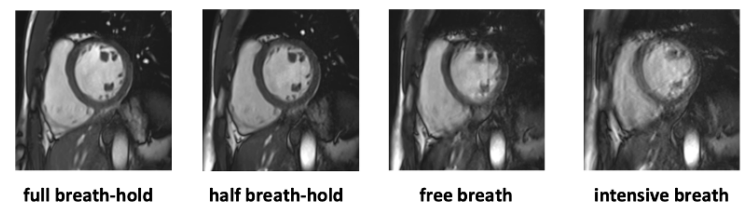}
    \label{breathings}
    \caption{The four different breathing intensities resulting in motion artifacts present in the CMRxMotion Challenge data (official challenge images).}
\end{figure}

\section{Method}
Our proposed method consists of: (1) resampling, preprocessing, and normalising the data, (2) loading a pretrained model and setting the appropriate training parameters, and (3) training the loaded network with data that have been augmented with a number of different deformation and intensity changes.

We refer to our 3D data as volumes and our short-axis slices as images. The data resampling, preprocessing and normalisation were done in 3D, while the data augmentation was performed over the slices.

Our method focuses on fine tuning an encoder during the segmentation task (with a segmentation head and decoder).

\subsection{Data resampling, preprocessing and normalisation}
First, we reoriented the volumes to the canonical orientation, which were then resampled. A crop was then applied to the region of interest. The data were subsequently split into training and validation subsets, where validation represented 20\% of the available subjects.

Second, the intensity was normalized using a histogram that we obtain out of the training samples. We perform the histogram standardization \cite{histstand} of all sets with the mentioned histogram.

\subsection{Architecture study}
In this work, we tested two different versions of the widely used U-Net architecture \cite{unet} to experiment if pretrained weights from a different problem and data domain could lead to improved results. The two approaches are:
\begin{itemize}
    \item a U-Net trained from scratch;
    \item a ResNet-based \cite{resnet} U-Net architecture with weights pretrained for ImageNet \cite{imagenet} classification.
\end{itemize}

\subsection{Data augmentation}
An important part of our study consisted of applying four different types of augmentation techniques, which have previously been shown to enhance the quality of the predictions on scans without the above-mentioned artifacts \cite{augmentations}. These augmentations were:
\begin{itemize}
    \item Random Motion: simulates movement in the tissue during the scan acquisition, and follows Shaw et al.~\cite{randommotion}.
    \item Random Ghosting: since respiratory motion is a cause for ``ghost" artifacts this augmentation simulates the effect along the phase-encoded direction, altering the intensity of the imaged structures.
    \item Random Bias Field: this field is modelled as a linear combination of polynomial functions, as in Van Leemput et al.~\cite{van1999automated}. These fields create inhomogeneity in the intensity of low frequencies throughout the image.
    \item Random Gamma: this intensity transform consists of a random change on the contrast of an image by raising its values.
\end{itemize}

\subsection{Cardiac MRI Dataset}
The dataset for our study was provided by the Extreme Cardiac MRI Analysis Challenge under Respiratory Motion. In particular, we employed the data from the segmentation challenge (task 2).

Training data represented a cohort of 20 subjects, which were scanned four different times, each under a different grade of respiratory intensity as in Figure \ref{breathings}, described as follows: (1) full breath hold, (2) half breath hold, (3) free breathing, and (4) intensive breathing. The evaluation data represented a cohort of five subjects with the same four different intensities of breathing.
For all subjects and breathing intensities, an expert radiologist recorded and labelled the end-diastolic (ED) and end-systolic (ES) phases.

\section{Experiment settings}
The experiment settings that were set during the training and inference of our method are described in the following section.

First, the architectures in Section 2.2 had the following details:
\begin{itemize}
    \item U-Net trained from scratch: 32 filters in the first out of five pairs of convolutional layers and a max-pooling layer after each of the four first pairs of convolutional layers. The convolutional layers used the ReLU activation and batch normalization. 
    \item U-Net pretrained: used a ResNet101 \cite{resnet} backbone as an encoder, pretrained with ImageNet \cite{imagenet}.
\end{itemize}

The general learning rate was set at $10^{-3}$ except for fine tuning, where the learning rate for the trained encoder was set at $10^{-4}$. The learning rate was scheduled to change on plateau with patience of 100 epochs, reducing the learning rate to half its previous value. The optimizer chosen was Adam.

The steps detailed in Sections 2.1 and 2.3 were done using the TorchIO library \cite{torchio}. The pretrained network was downloaded from PyTorch Segmentation Models~\cite{pytorch}.

The augmentation policy that was applied consisted of applying always one of the techniques described in section 2.3, were the random motion had three times the chances to be applied.

\section{Results}
We present our results in two different sections: (1) validation results, and (2) evaluation results which correspond to the results provided by the challenge platform.

\subsection{Validation results}
Table \ref{tab:train_results} shows the results on the validation split of the training set. Four different models are listed, indicating whether the training included augmentations, and if the weights were trained from scratch or pretrained on ImageNet.

\begin{table}[ht!]
\centering
\begin{tabular}{l@{\hskip 2em}cccc}
\toprule
                              & \multicolumn{4}{c}{\textbf{DICE}}                       \\ \midrule
                              & \textbf{LV} & \textbf{MYO} & \textbf{RV} & \textbf{ALL} \\ \midrule
\textbf{U-Net (scratch)}      & 0.97        & 0.945        & 0.963       & 0.959        \\
\textbf{U-Net (scratch) Augs} & 0.974       & 0.949        & 0.965       & 0.963        \\ \midrule
\textbf{U-Net (ImageNet)}     & 0.97        & 0.948        & 0.966       & 0.962        \\
\textbf{U-Net (ImageNet) Augs} & \textbf{0.976} & \textbf{0.952} & \textbf{0.97} & \textbf{0.966} \\ \bottomrule
\end{tabular}
\vspace{2mm}
\caption{Validation results (DICE). Augs indicates the additional data augmentation. Best results in bold.}
\label{tab:train_results}
\end{table}

\subsection{Evaluation results}
Table \ref{tab:eval_results} shows the results of the inference of the evaluation data on the challenge platform. The four models are the same used in the previous section.
\begin{table}[ht!]
\centering
\begin{tabular}{@{}l@{\hskip 2em}cccccc@{}}
\toprule
                               & \multicolumn{3}{c}{\textbf{DICE}}        & \multicolumn{3}{c}{\textbf{Hausdorff (mm)}}   \\ \midrule
                               & \textbf{LV} & \textbf{MYO} & \textbf{RV} & \textbf{LV} & \textbf{MYO} & \textbf{RV} \\ \midrule
\textbf{U-Net (scratch)}       & 0.88        & 0.768        & 0.789       & 11.78       & 7.64         & 11.37       \\
\textbf{U-Net (scratch) Augs}  & 0.88        & 0.771        & 0.782       & \textbf{10.57}       & 7.74         & 11.87       \\ \midrule
\textbf{U-Net (ImageNet)}      & 0.879       & 0.796        & 0.826       & 11.4        & 6.2          & 8.71        \\
\textbf{U-Net (ImageNet) Augs} & \textbf{0.883}       & \textbf{0.797}        & \textbf{0.851}       & 11.04       & \textbf{5.64}         & \textbf{7.77}        \\ \bottomrule
\end{tabular}
\vspace{2mm}
\caption{Evaluation results (DICE and Hausdorff 95). Augs indicates the additional data augmentation. Best results in bold.}
\label{tab:eval_results}
\end{table}

\section{Conclusions}
In this paper, we proposed starting the training from weights that were obtained in a classification problem in another data domain. In addition, we proposed an augmentation policy consisting of four different augmentations with random motion being applied three times more than the rest.

From the quantitative analysis that was done in the validation split of the training data, and the evaluation data from the official platform, we can say that both additions corresponded to an increase of the quality of segmentation. Moreover, the adoption of the pretrained weights also accelerated training times.

\section*{Acknowledgment}
This publication has emanated from research conducted with the financial support of Science Foundation Ireland under Grant number 18/CRT/6183.

%
%
%
%

\bibliographystyle{splncs04}
\bibliography{bibliography}

\end{document}